\documentclass[preprint,preprintnumbers,showpacs,amsmath,amssymb]{revtex4-1}

\usepackage{graphicx}
\usepackage{dcolumn}
\usepackage{bm}

\begin{document}


\title{Energy dependence of the inelasticity in
$pp/p\overline{p}$ collisions from experimental information on charged particle multiplicity distributions}

\author{P.C. Beggio$^{a,1}$ and F.R. Coriolano$^{b,1}$}


\affiliation{
$^{a}$Laborat\'orio de Ci\^encias Matem\'aticas -
LCMAT. \\
$^{b}$Programa de P\'os Gradua\c{c}\~{a}o em Ci\^encias Naturais - PPGCN. \\
$^{1}$Universidade Estadual do Norte Fluminense Darcy Ribeiro -
UENF, 28013-602, Campos dos Goytacazes, RJ, Brazil.}


\begin{abstract}

The dependence of the inelasticity in terms of the center of mass
energy is studied in the eikonal formalism, which provides
connection between elastic and inelastic channels. Due to the
absence of inelasticity experimental datasets, the present analysis
is based on experimental information available on the full phase
space multiplicity distribution covering a large range of energy,
namely 30 $<$ $\sqrt{s}$ $\leq$ 1800 GeV. Our results indicate that
the decrease of inelasticity is consequence of minijets production
from semihard interactions arising from the scattering of gluons
carrying only a very small fractions of the momenta from their
parent protons. Alternative methods of estimating the inelasticity
are discussed and predictions to the LHC energies are presented.

\end{abstract}



\maketitle

\section{Introduction}

In $p+p(\overline{p})$ collisions at center of mass energy,
$\sqrt{s}$, the effective energy left behind by the two leading
protons, or correspondingly the inelasticity $K$
\cite{Fiete,Braz_Japn,YH,Navarra}, is an essential concept because
it defines the energy effectively used for producing $n$ new
secondary particles. That in turn, determines the dynamics of the
interaction in high-energy hadronic and nuclear collisions. The
inelasticity varies from event to event, so that one has to
introduce an inelasticity distribution $\chi(K,s)$ normalized by
\cite{Fowler1984}
\begin{equation}
\int_{0} ^{1} \chi(K,s)\, dK\,=\,1\,. \label{eq1}
\end{equation}
Experimental data on $K$ are very limited and the form of its
distribution function, $\chi(K,s)$, has not yet been stablished. It
is known as the only experimental information available on
$\chi(K,s)$ is from $pp$ interactions at $\sqrt{s}$=16.5 GeV, which
exhibits a maximum at $\sim$ $0.5$ \cite{Brick1981}. At the ISR
energies the mean inelasticity is approximately constant with $<K>$
$\sim$ $0.5$ \cite{Kadija}.

The energy dependence of the inelasticity is an important problem
which has been subject of discussions
\cite{Fiete,Wibig,Navarra2003,Navarra1993,Navarra1994,Musulmanbekov,BeggioNPA2011}.
As example, comparing $p+p(\overline{p})$ with $e^{-}e^{+}$
collisions the $\sqrt{s}$ dependence of the inelasticity in
$p+p(\overline{p})$ collisions was calculated in \cite{Fiete} for
three different assumptions on the parameters involved in the
analysis and the results were compared with the theoretical study
from \cite{Kadija}.

Although, as mentioned, experimental information on $K(s)$ is
limited, the probability for producing $n$ charged particles in
final states $P_{n}(s)$, or simply multiplicity distribution, is
strictly connected with the inelasticity concept
\cite{Musulmanbekov,BeggioNPA2011}. Thus, we can study $P_{n}(s)$
features in order to derive informations on the $K(s)$ behavior,
since there are experimental informations available in the full
space phase for $P_{n}(s)$ covering the interval of 30 $<$
$\sqrt{s}$ $\leq$ 1800 GeV \cite{ABC,UA51,Alexopoulos}.

With that in mind, we have studied the relation between $P_{n}(s)$
and $K(s)$ in the framework of a \emph{phenomenological procedure}
related to $P_{n}(s)$ \cite{Lam 1982,BeggioMV}, as well as a formula
connecting the inelasticity to the imaginary part of the eikonal
function in the impact parameter $b$ space, $\chi_{I}(s,b)$, which
was obtained in \cite{BeggioNPA2011}. However, in the analysis done
in \cite{BeggioNPA2011} the $P_{n}(s)$ dataset studied was restrict
to collision energies of 52.6, 200, 546 and 900 GeV, and only a
limited success was reached in describing the $P_{n}(s)$ data at 200
and 900 GeV. Here, however, we treat the full phase space $P_{n}(s)$
and $K(s)$ at $\sqrt{s}=$ 30.4, 44.5, 52.6, 62.2, 300, 546, 1000 and
1800 GeV) \cite{ABC,UA51,Alexopoulos}.

Since in our studies $K(s,b)\,\propto\,\chi_{I}(s,b)$, in
\cite{Beggio Luna} we have updated the eikonal formalism of the
aforementioned \emph{phenomenological procedure} in order to
describe, in a connected way, $p+p(\overline{p})$ observables in
both elastic and inelastic channels through the unitarity condition
of the S-Matrix in impact parameter space. All the parameters of the
eikonal function, $\chi_{pp}^{\overline{p}p}(s,b)$, were determined
carrying out a global fit to all high energy forward $pp$ and
$\overline{p}p$ scattering data above $\sqrt{s}$=10 GeV, namely the
total cross section, $\sigma_{tot}^{pp,\overline{p}p}$, the ratio of
the real to imaginary part of the forward scattering amplitude
$\rho^{pp,\overline{p}p}$, the elastic differential scattering cross
sections $d\sigma_{el}^{\overline{p}p}/dt$ at $\sqrt{s}$=546 GeV and
$\sqrt{s}$=1.8 TeV as well as the TOTEM datum on $\sigma_{tot}^{pp}$
at $\sqrt{s}$=7 TeV. The results obtained in \cite{Beggio Luna} were
compared with the correspondent experimental information and also
with the full phase space $P_{n}$ and the $H_{q}$ moments, yielding
successful descriptions of all experimental data. In \cite{Beggio
JPhys G 2017} the \emph{phenomenological procedure} from
\cite{Beggio Luna} was applied to investigate the $\sqrt{s}$
dependence of the parton-parton inelastic cross sections,
parton-parton inelastic overlap functions and the $C_{q}$ moments in
proton interactions from $\sqrt{s}$=10 to 14000 GeV, providing also
predictions for the ratio $\sigma_{el}(s)/\sigma_{tot}(s)$ as a
function of the $\sqrt{s}$, in agreement with the experimental data.
Therefore, the success in that global description of elastic and
inelastic hadronic observables, over wide interval of $\sqrt{s}$
\cite{Beggio Luna,Beggio JPhys G 2017}, motivated us to investigate
the problem of the $\sqrt{s}$ dependence of the $K(s)$ from
$P_{n}(s)$ studies.

The main purpose of this paper is to apply the
\emph{phenomenological procedure} formalism in full phase space
$P_{n}(s)$ from \cite{Beggio Luna}, also applied in \cite{Beggio
JPhys G 2017}, in order to study the energy dependence of the
inelasticity based on the experimental information from
$p+p(\overline{p})$ multiplicity distributions, since experimental
data on $K(s)$ are very limited.

The paper is organized as follows: in the next section we discuss
the main ideas associated with the \emph{phenomenological procedure}
as well as their inputs. In Section III, we apply the theoretical
formalism computing the inelasticity as a function of $b$ at fixed
$\sqrt{s}$, discussing the results. Inelasticity predictions to the
LHC energies are made. The concluding remarks are the content of the
Section IV.

\section{\emph{Phenomenological procedure}}

\subsection{The $P_{n}(s)$ model}

The multiplicity distribution is defined at $\sqrt{s}$ in terms of
the topological cross section, $\sigma_{n}$, and the inelastic cross
section, $\sigma_{in}$, by the formula
\begin{equation}
P_{n}(s)=\frac{\sigma_{n}(s)}{\sum_{n}\,
\sigma_{n}(s)}=\frac{\sigma_{n}(s)}{\sigma_{in}(s)}\,. \label{eq2}
\end{equation}
In the impact parameter formalism a normalized $P_{n}(s)$ may be
constructed by summing contributions coming from $p+p(\overline{p})$
collisions taking place at fixed $b$ and $\sqrt{s}$. Thus $P_{n}(s)$
is written as
\begin{equation}
P_{n}(s)=\frac{\sigma_{n}(s)}{\sigma_{in}(s)}=\frac{ \int
d^2b\,[1-e^{-2\,\chi_{I}(s,b)}][\frac{\sigma_{n}(s,b)}{\sigma_{in}(s,b)}]}{\int
d^2b\,[1-e^{-2\,\chi_{I}(s,b)}]} \label{eq3}
\end{equation}
where the $\sigma_{n}(s)$ is decomposed into contributions from each
impact parameter $b$, and
$\sigma_{in(s,b)}=G_{in(s,b)}=[1-e^{-2\,\chi_{I}(s,b)}]$ is the
weight function, called inelastic overlap function. As in its
original formulation \cite{Lam 1982,BeggioMV} the quantity in
brackets scales in KNO sense and we can rewritten the last Eq. as
\begin{equation}
P_{n}(s)=\frac{ \int d^2b\,\frac{[1-e^{-2\,\chi_{I}(s,b)}]}{\langle
n(s,b) \rangle} [\langle n(s,b)
\rangle\,\frac{\sigma_{n}(s,b)}{\sigma_{in}(s,b)}]}{\int
d^2b\,[1-e^{-2\,\chi_{I}(s,b)}]}\,, \label{eq4}
\end{equation}
where $\langle n(s,b) \rangle$ is the average number of particles
produced at $b$ and $\sqrt{s}$ and its factorizes as \cite{BeggioMV}
\begin{equation}
\langle n(s,b) \rangle=\langle N(s) \rangle\,f(s,b)\,. \label{eq5}
\end{equation}
In this equation $\langle N(s) \rangle$ is the average multiplicity
at $\sqrt{s}$ and $f(s,b)$ is called multiplicity function.
Similarly to KNO, it is introduced the elementary multiplicity
distribution related to microscopic processes
\begin{equation}
\psi\left(\frac{n}{\langle n(s,b) \rangle}\right)=\langle n(s,b)
\rangle\frac{\sigma_{n}(s,b)}{\sigma_{in}(s,b)}\,. \label{eq6}
\end{equation}
As in previous works \cite{BeggioNPA2011,BeggioMV,Beggio Luna,Beggio
JPhys G 2017,BeggioHama,BeggioBJP,BeggioNPA2013}, we have assumed
that the particles created at $\sqrt{s}$ and $b$ follows the KNO
form of the Negative Binomial distribution, or Gamma distribution,
normalized to 2
\begin{equation}
\psi\left(\frac{n}{\langle n(s,b)
\rangle}\right)=2\,\frac{k^k}{\Gamma(k)}\left[\frac{n}{\langle
n(s,b) \rangle}\right]^{k-1}e^{-k\left[\frac{n}{\langle n(s,b)
\rangle }\right]}\,
 \label{eq7}
\end{equation}
which is characterized by the $k$ parameter and $\Gamma$ represents
the usual gamma function. Its choose was motivated by the fact that
this distribution arises as the dominant part of the solution of the
equation for three gluon branching process in the very large $n$
limit \cite{durand001}. This branching equation, which takes into
account only gluon bremsstrahlung process, gives the main
contribution at high energies since semihard gluons dominate the
parton-parton cross sections. Thus, with the Eqs. (\ref{eq5}) and
(\ref{eq6}), the Eq. (\ref{eq4}) becomes
\begin{equation}
P_{n}(s)=\frac{ \int
d^2b\,\frac{[1-e^{-2\,\chi_{I}(s,b)}]}{f(s,b)}[\psi\,(\frac{n}{\langle
N(s) \rangle\,f(s,b)})]} {\langle N(s) \rangle\,\int
d^2b\,[1-e^{-2\,\chi_{I}(s,b)}]}\,. \label{eq8}
\end{equation}
Now, to define $f(s,b)$ in terms of the imaginary eikonal
$\chi_{I}(s,b)$ we have assumed that
\begin{enumerate}
\item the fraction of $\sqrt{s}$, which is deposited by the two leading
protons for particle production in a collision at $b$, represented
by $\sqrt{s'}$, is proportional to $\chi_{I}(s,b)$:
\begin{equation}
\sqrt{s^{'}}=\beta(s)\,\chi_{I}(s,b)\,,
 \label{eq9}
\end{equation}
where $\beta(s)$ is a function to be defined.
\item The average number of produced particles
depends on the $\sqrt{s^{'}}$ at each $b$ value in a power law form
\begin{equation}
\langle n(s,b)
\rangle=\gamma\,\left(\frac{s^{'}}{s^{'}_{0}}\right)^{\zeta(s)},
\label{eq10}
\end{equation}
\end{enumerate}
where $s^{'}_{0}$=1 GeV$^{2}$. Substituting the Eq. (\ref{eq9}) into
(\ref{eq10}) we obtain the energy and impact parameter dependence of
$\langle n \rangle$
\begin{equation}
\langle n(s,b) \rangle=\frac{\gamma\,[\beta(s) \,
\chi_{I}(s,b)]^{2\,\zeta(s)}}{(s_{0}^{'})^{\zeta(s)}}\,.
\label{eq10a}
\end{equation}
The $\gamma$ parameter and the $\zeta(s)$ function will be discussed
in the next subsection.

The physical motivation of the Eq. (\ref{eq9}) is that the eikonal
may be interpreted as an overlap, on the impact parameter plane, of
two colliding matter distributions \cite{Barshay}. Physically, the
Eq. (\ref{eq9}) corresponds to the effective energy for particle
production, then we can write $\sqrt{s^{^{'}}} \equiv E_{eff}$.

The Eq. (\ref{eq10}) deserves a more detailed comment: a power law
dependence of the multiplicity on the energy emerged in the context
of statistical and hydrodynamical models. It also was successfully
applied in the context of the parton model, either connecting KNO
and Bjorken scaling or treating the violation of the KNO scaling and
can also arise from a simple picture of branching decay producing a
tree structure (see \cite{BeggioMV} and references therein). In
\cite{Troshin} the authors reproduced the power like energy behavior
of the mean multiplicity in the hadronic multiparticle production
model with antishadowing, which provided estimated values of the
average multiplicity over a large energy interval, in good agreement
with the data and predicting multiplicities at the LHC energies.
Based on the gluon saturation scenario (Color Glass Condensate
approach), in \cite{Levin}, the authors showed that the power law
energy dependence of charged hadron multiplicity leads to a very
good description of the LHC experimental data in both, $pp$
($s^{0.11}$) and AA (nucleus-nucleus) ($s^{0.145}$) collisions,
including the ALICE data in Pb+Pb collisions at 2.76 TeV and showed
that this different energy dependence can be explained by inclusion
of a strong angular-ordering in the gluon decay cascade. A power law
behavior is characteristic of several analyses of experimental data
on hadronic interactions and also several theoretical approaches.
Thus, at the present stage of our studies, the power law for the
multiplicity seems a hypothesis reasonable.

Matching the Eqs. (\ref{eq5}), (\ref{eq9}) and (\ref{eq10}) we have
\begin{equation}
f(s,b)=\frac{\gamma}{\langle
N(s)\rangle}\left[\frac{\beta(s)}{\sqrt{s^{'}_{0}}}\right]^{2\,\zeta(s)}[\chi_{I}(s,b)]^{2\,\zeta(s)}\,
\label{eq11}
\end{equation}
and defining $\xi(s)$ in the last Eq. as
\begin{equation}
\xi(s)\equiv\frac{\gamma}{\langle
N(s)\rangle}\left[\frac{\beta(s)}{\sqrt{s^{'}_{0}}}\right]^{2\,\zeta(s)}\,
\label{eq12}
\end{equation}
the Eq. (\ref{eq11}) can be written as
\begin{equation}
f(s,b)=\xi(s)[\chi_{I}(s,b)]^{2\,\zeta(s)}. \label{eq13}
\end{equation}
In turn, substituting the Eq. (\ref{eq13}) into Eq. (\ref{eq8})
results
\begin{equation}
P_{n}(s)=\frac{ \int
d^2b\,\frac{[1-e^{-2\,\chi_{I}(s,b)}]}{\xi(s)[\chi_{I}(s,b)]^{2\,\zeta(s)}}[\psi\,(\frac{n}{\langle
N(s) \rangle\,\xi(s)[\chi_{I}(s,b)]^{2\,\zeta(s)}})]} {\langle N(s)
\rangle\,\int d^2b\,[1-e^{-2\,\chi_{I}(s,b)}]}\,, \label{eq14}
\end{equation}
with $\xi(s)$ determined by the usual normalization conditions on
the charged $P_{n}(s)$ ($\int P_{n}\,dn\,=\,\int\,
P_{n}\,n\,dn\,=2$), explicitly we have obtained \cite{BeggioMV}
\begin{equation}
 \xi(s)=\frac
  {\int d^2 b\,[1-e^{-2\,\chi_{I}(s,b)}]}
  {\int d^2 b\,[1-e^{-2\,\chi_{I}(s,b)}]\,[\chi_{I}(s,b)]^{2\,\zeta(s)}}.\
 \label{eq15}
\end{equation}
The formalism permits the calculation of the $P_{n}(s)$, Eq.
(\ref{eq14}), once an eikonal parametrization is assumed and
appropriate values to the parameters $k$ and $\zeta(s)$ are adjusted
in order to provide reliable results concerning calculations of
strongly interacting processes, as discussed in next subsection.

The physical picture of the $P_{n}(s)$ is discussed in detail in
\cite{Beggio Luna} and asserts that the full phase space $P_{n}(s)$
is constructed by summing contributions from parton-parton
collisions occuring at each value of $b$, with the formation of
strings that subsequently fragments into hadrons. The idea of string
formation for multiparticle production is similar to the Lund model
\cite{Lund model}.

\subsection{QCD-inspired eikonal model, $k$, $\zeta(s)$ and $\gamma$ parameters}

We adopted the QCD-inspired eikonal model referred as Dynamical
Gluon Mass (DGM) model \cite{luna008}, which incorporates soft and
semihard processes using a formulation compatible with analycity and
unitarity principles. The eikonal function is written in terms of
even and odd eikonal parts, connected by crossing symmetry and this
combination leads \cite{luna008,luna009}:
\begin{equation}
\chi_{pp}^{\overline{p}p}{(s,b)}=\chi^{+}{(s,b)}\pm\chi^{-}{(s,b)}\,.
\label{eq16}
\end{equation}
The even eikonal is written as the sum of quark-quark, quark-gluon
and gluon-gluon contributions
\begin{equation}
\chi^{+}{(s,b)}=\chi_{qq}{(s,b)}+\chi_{qg}{(s,b)}+\chi_{gg}{(s,b)}\,.
 \label{eq17}
\end{equation}
\begin{equation}
\chi^{+}{(s,b)}=i[\sigma_{qq}{(s)\,W(b;\mu_{qq})}+\sigma_{qg}{(s)\,W(b;\mu_{qg})}+
\sigma_{gg}{(s)\,W(b;\mu_{gg})}]\,.
 \label{eq18}
\end{equation}
where $W(b;\mu_{ij})=\mu^{5}b^{3}K_{3}(\mu_{ij}\,b)/96\pi$ is the
overlap density for the partons at $b$, $\sigma_{ij}(s)$ are the
elementary subprocess cross sections of colliding quarks and gluons
($i,j=q,g$) and $K_{3}(x)$ is the modified Bessel function of second
kind. The eikonal functions $\chi_{qq}{(s,b)}$ and
$\chi_{qg}{(s,b)}$ are needed to describe the lower energy forward
data and are parametrized with inputs from Regge phenomenology (for
details see \cite{luna008}).

It is important to note that the term $\chi_{gg}{(s,b)}$ gives the
main contribution to the asymptotic behavior of the
$p+p(\overline{p})$ total cross sections and its energy dependence
comes from gluon-gluon cross section
\begin{equation}
\sigma_{gg}(s)=C_{gg}(s)\,\int_{4m_{g}^{2}/s}^{1}d\tau\,F_{gg}(\tau)\,\widehat{\sigma}(\widehat{s}),
\label{eq19}
\end{equation}
where $\tau=x_{1}x_{2}=\widehat{s}/s$,
$F_{gg}(\tau)=\int_{\tau}^{1}\frac{dx}{x}g(x)g(\frac{\tau}{x})$ is
the convoluted structure function for a pair gluon-gluon,
$\widehat{\sigma}(\widehat{s})$ is the total cross section for the
subprocess $gg \rightarrow gg$ and $C_{gg}$ is a free parameter
\cite{Beggio Luna,Beggio JPhys G 2017}.

Relating to the term $\chi^{-}(s,b)$, Eq. (\ref{eq16}), the role of
the odd eikonal is to account the difference between $pp$ and
$p\overline{p}$ channels at low energies and it is written as
\begin{equation}
\chi^{-}{(s,b)}=C^{-}\sum\frac{m_{g}}{\sqrt{s}}e^{i\pi/4}W(b;\mu^{-}),
 \label{20}
\end{equation}
where $m_{g}=364$ $\pm\,26$ MeV is an infrared mass scale
\cite{luna010} and $C^{-}$ a fitted constant. All the DGM model
parameters used in this work were determined in \cite{Beggio Luna}
carrying out a global fit to all high energy forward
$p+p(\overline{p})$ scattering data above $\sqrt{s}$ $=$ 10 GeV,
namely the total cross section, $\sigma_{tot}^{pp,p\overline{p}}$,
the ratio of the real to imaginary part of the forward scattering
amplitude, $\rho^{pp,p\overline{p}}$, the elastic differential
scattering cross sections, $d\sigma^{p\overline{p}}/dt$, at
$\sqrt{s}$ $=$ 546 GeV and $\sqrt{s}$ $=$ 1.8 TeV as well as the
TOTEM datum on $\sigma_{tot}^{pp}$ at 7 TeV. The $\chi^{2}/DOF$ for
the global fit was 0.98 for 320 degrees of freedom. The values of
the fitted parameters and the results of the fits to
$\sigma_{tot}^{pp,p\overline{p}}$, $\rho^{pp,p\overline{p}}$ and
$d\sigma^{p\overline{p}}/dt$ are presented and discussed in
\cite{Beggio Luna}. Thus, all free parameters of the DGM model were
completely determined from elastic channel fits.

Now, we see from Eqs. (\ref{eq14}) and (\ref{eq7}) that the only
free parameters in the $P_{n}(s)$ analysis are $k$ and $\zeta(s)$.
With respect to $k$, assuming the Gamma distribution, Eq.
(\ref{eq7}), experimental data on $e^{+}e^{-}$ annihilation were
fitted obtaining $k=10.775\,\pm\,0.064$ ($\chi^{2}/N_{DF}=2.61$)
\cite{BeggioMV}. By using the DGM eikonal model parametrization,
fixing the value of $k=10.775$ and assuming $\zeta(s)$ as the single
fitting parameter, $p+p(\overline{p})$ full phase space $P_{n}(s)$
experimental data in the interval 30.4 GeV $\leq$ $\sqrt{s}$ $\leq$
1800 GeV were fitted by the Eq. (\ref{eq14}) \cite{Beggio Luna},
yielding the $\zeta(s)$ values summarized in Table I, together with
the values of $\xi(s)$ computed from Eq. (\ref{eq15}). The $\langle
N(s) \rangle$ values were obtained from experimental data
\cite{ABC,UA51,Alexopoulos}, Table I. The $\zeta(s)$ energy
dependence can be described in a consistent way through the function
\cite{Beggio Luna}
\begin{equation}
\zeta(s)=0.189+0.00197\,[ln(s)]^{1.536}\,. \label{eq21}
\end{equation}
This procedure in fact does provided an excellent description of the
$P_{n}(s)$ data at high multiplicities, avoiding the introduction of
more free parameters. The $P_{n}(s)$ plots from Ref. \cite{Beggio
Luna} are reproduced in this work, as shown at the top panels in
Figs. 1 to 8. All the $P_{n}(s)$ results are in good agreement with
the experimental points \cite{ABC,UA51,Alexopoulos}, the values of
$\chi^{2}/N_{DF}$ are presented in Table I. Theoretical predictions
in full phase space $P_{n}(s)$ at the LHC energies of $\sqrt{s}$=7
and 14 TeV are shown in Fig. 9.

With respect to $\gamma$ parameter, it is unnecessary to calculate
the $P_{n}(s)$ since it is absorbed into the definition of the
normalization condition $\xi(s)$, Eq. (\ref{eq12}) and, in turn,
$\xi(s)$ is calculated by Eq. (\ref{eq15}). However, we cannot
calculate $K(s,b)$ until its values are known (see Eq. (\ref{eq27})
bellow) and, in this formalism, we cannot estimate the $\gamma$
value directly from $p+p(\overline{p})$ data. This parameter was
introduced in the $P_{n}(s)$ \emph{phenomenological procedure} by
Eq. (\ref{eq10}) on the hypothesis that the average number of
produced particles depends on the effective energy for particle
production through a power law. In order to have a reliable estimate
of $\gamma$, from a strongly interacting system, we considered the
experimental data on $e^{-}e^{+}$ annihilation as a possible source
of information concerning parton-parton interaction in
$p+p(\overline{p})$ collisions and adopted the results from Ref.
\cite{BeggioHama}, where average multiplicity data in $e^{+}e^{-}$
annihilations, covering the interval 10 $\leq$
$(\sqrt{s})_{e^{+}e^{-}}$ $\leq$ 200 GeV, were fitted by Eq. (10),
yielding the values of $\gamma$=3.36 and
$\zeta_{(e^{+}e^{-})}=0.200$, with $\chi^{2}/N_{DF}$=0.94. In
$e^{-}e^{+}$ annihilation probably one $q\overline{q}$ pair has
triggered the multitude of the final particles and, despite the fact
that in $p+p(\overline{p})$ more channels should contribute, this
approximation seems reasonable because when the average multiplicity
increases, the relevance of the original parton may decreases
\cite{BeggioMV}.

It is important to note that the impact parameter dependence of the
inelasticity for some collision energies studied in this paper also
was studied in \cite{BeggioNPA2011}, where the obtained inelasticity
values are much larger than the values found in this work. The
different values assigned to the gama parameter in the Eq.
(\ref{eq27}), in each analysis, is the main reason for this
difference. In \cite{BeggioNPA2011} it was used the value
$\gamma$=2.09 obtained in \cite{BeggioMV} where average multiplicity
data in $e^{-}e^{+}$ annihilations, in the interval 5.1 $\leq$
$(\sqrt{s})_{e^{+}e^{-}}$ $\leq$ 183 GeV, were fitted by Eq. (10)
giving $\gamma$=2.09 and $\zeta_{(e^{+}e^{-})}=0.258$ with
$\chi^{2}/N_{DF}$=8.89. As explained before, here we have adopted
$\gamma$=3.36 in reason of a better $\chi^{2}/N_{DF}$ value than
those obtained from $\gamma$=2.09. At an example level, at
$\sqrt{s}$=52.6 GeV and $b$=0 the corresponding values of the
parameters are $\xi(s)$=1.639, $\langle N(s)\rangle$=11.55,
$\zeta(s)$=0.239, $\chi_{I}(s,0)$=1.305 and $\gamma$=3.36. By using
them in the Eq.(\ref{eq27}) result $K(s,0)$$\approx$0.48 . By
changing only the value of $\gamma$ to 2.09 we obtain
$K(s,0)$$\approx$1.25, which is clearly wrong.

\section{Energy dependence of the inelasticity and discussions}

In $p+p(\overline{p})$ collisions at $\sqrt{s}$ the effective energy
for particle production, $E_{eff}$, is the energy left behind by two
leading protons and, using four-vector, it may be written
\cite{Kadija}
\begin{equation}
E_{eff}=\sqrt{s}-(E_{leading,1}+E_{leading,2})\,, \label{eq22}
\end{equation}
or
\begin{equation}
E_{eff}=\sqrt{s} - 2\,E_{leading}\,,\label{eq23}
\end{equation}
in the case of symmetric events \cite{Fiete} \cite{Kadija} and, for
quantitative estimation of the inelasticity, we have used the
definition \cite{Navarra}
\begin{equation}
K=E_{eff}/\sqrt{s}\,, \label{eq24}
\end{equation}
($0 \leq K \leq 1$). We see from Eq. (\ref{eq9}) that
$\sqrt{s'}=E_{eff}$ \cite{BeggioNPA2011}, and hence we can rewrite
the last Eq. in the form
\begin{equation}
K(s,b)=\, \frac{\beta(s)\,
\chi_{I}(s,b)}{2\,\sqrt{s}}\,.\label{eq25}
\end{equation}
The factor 2 is due the fact that the $P_{n}(s)$ data are normalized
to 2. In turn, the $\beta(s)$ function is related with $\xi(s)$ by
Eq. (\ref{eq12}), explicitly we have
\begin{equation}
\beta(s)=\left[\frac{\xi(s)\,\langle N(s) \rangle
(s_{o}^{'})^{\zeta(s)}}{\gamma}\right]^{\frac{1}{2\,\zeta(s)}}\,.
\label{eq26}
\end{equation}
Using the Eq. (\ref{eq26}) we can rewritten the Eq. (\ref{eq25}) in
the form
\begin{equation}
K(s,b)=\left[\frac{\xi(s)\,\langle N(s) \rangle
(s_{o}^{'})^{\zeta(s)}}{\gamma}\right]^{\frac{1}{2\,\zeta(s)}}\,\frac{\chi_{I}(s,b)}{2\,\sqrt{s}}\,.
\label{eq27}
\end{equation}
With respect to last expression, the DMG eikonal function
$\chi_{I}(s,b)$ is completely determined from only elastic channel
data analysis (subsection II.B), $\xi(s)$ is determined by the
normalized condition given by the Eq. (\ref{eq15}), the $\langle
N(s) \rangle$ values are obtained from experiments and the
$\zeta(s)$ values were obtained by full phase space $P_{n}(s)$ fits
\cite{Beggio Luna} and parametrized by the Eq. (\ref{eq21}). Thus,
by fixing the value of $\gamma=3.36$, as discussed in the Subsection
II.B, we have calculated the $K(s,b)$ as a function of the impact
parameter $b$ and the results are displayed in Figs. 1 to 8. The
inelasticity behavior is essentially the same at the energies of
$\sqrt{s}$ = 62.2 and 44.5 GeV, Fig. 4. The same occurs at 1000 and
546 GeV, Fig. 7. It seems consequence from the fact that at 62.2 and
1000 GeV the theoretical $P_{n}$ does not fits satisfactorily the
experimental points in the tail of the distributions.

At the ISR energies the average inelasticity is determined to be
about 0.5 \cite{Kadija,Golyak}. Interestingly in the present
analysis is that the average inelasticity at ISR, when $b$ $\sim$ 0,
yields the same value, specifically: $<K>_{ISR}$ = (0.54 + 0.49 +
0.48 + 0.50)/4 $\sim$ 0.5, however, the choice of $b$ $\sim$ 0 is so
arbitrary. Based on the results displayed in Fig. 4 and by using the
formulae of mathematical expectation of the function $K(s,b)$, we
have calculated the average impact parameter, $<b>$, at each ISR
energy and the corresponding $K(s,<b>)$ values as well as the new
value of $<K>_{ISR}$ $\sim$ 0.16. The results are summarized in
Table II and the average inelasticity, thus obtained, do not agree
with those from \cite{Kadija,Golyak}. However, we recall that the
impact parameter dependence of the inelasticity was not analysed in
the framework of the both mentioned works, \cite{Kadija,Golyak}.

From Fig. 9, where the plots of $K(s)$ versus $b$ at the energies
investigated in this work are presented together, is possible see
that the particle production processes tend to be more peripheral
($b > 1$ $fm$) at the ISR energies of $30.4$, $44.5$, $52.6$ and
$62.2$ GeV when compared with the results from other energies
investigated. In this interval of $b$ the $K(s)$ values, at the ISR,
are rather greater than $K(s)$ values at the others energies at
fixed value of $b$. In order to substantiate this statement in Fig.
10 we show the ratios $K(s,b)/K(30.4,0)$ calculated for different
collision energies at the impact parameter values of $1.0$, $1.1$,
$1.25$ and $1.5$ $fermi$. Based in Fig. 1 we have used
$K(30.4,0)$=0.54. In fact, the results presented in Fig. 10 are
indicatives that the particle production is more peripheral at the
mentioned ISR energies than at other energies studied. This behavior
of $K(s)$ is compatible with the minijets production, since semihard
processes are more central in the impact parameter than purely soft
processes and do not use much collision energy \cite{DDeus}. The
inelasticity $K$ is proportional to the $\chi_{I}(s,b)$, Eq.
(\ref{eq27}), and in the DGM eikonal model the gluon semihard
contribution $\chi_{gg}(s,b)$, Eq. (\ref{eq17}), dominates at high
energy and the rise of the cross sections with $\sqrt{s}$ is
consequence of the increasing number of soft gluons populating the
colliding particles, increasing, therefore, the probability of
perturbative gluon-gluon collisions at small $x$, which can leads to
the appearance of minijets and, as mentioned, do not use much
collision energy. This scenario leads to the conclusion that the
$K(s)$ decreases as a consequence of the minijet production from
semihard soft gluon-gluon interactions when $\sqrt{s}$ increases.

We show in Fig. 11 the energy dependence of the $K(s)$ calculated at
$b$ $\approx$ $0$, Eq. (\ref{eq27}), and observe a marked decrease
in the inelasticity from ISR to LHC, while at the $\sqrt{s}$ $>$ 7
TeV the inelasticity shows a slow decrease. The error bars represent
the uncertainties of the parameters $\gamma$ and $\zeta$ propagated
to the inelasticity values. The star symbol represents theoretical
predictions at the LHC and the solid line is drawn only as guidance
for the points. The LHC has measured the multiplicity distributions
in a limited pseudorapidity range
\cite{CMS2011,LHCb2012,LHCb2014,ALICE2016,ALICE2017,Fiete2017cap},
and for this reason we do not compare our results with those from
LHC.

We observe that the structure found around the peak in the
$P_{n}(s)$ data at higher energies, which appears in the region of
low multiplicities, has not been considered in the analysis done in
\cite{Beggio Luna}. However, the $P_{n}(s)$ approach used describes
very well the energy dependence of the $F$-moments and reproduces
the $H_{q}$ versus $q$ oscillations observed in the experimental
data and predicted by QCD \cite{Beggio Luna,BeggioNPA2013}.

With respect to alternative methods of estimating the inelasticity,
in \cite{Fiete}, the coefficient of inelasticity in
$p+p(\overline{p})$ collisions and its possible $\sqrt{s}$
dependence was estimated by comparing $p+p(\overline{p})$ with
$e^{-}e^{+}$ collisions for three different assumptions on the
values for both the parameters involved in the analysis, namely
$n_{0}$ and $\Delta m$. The parameter $n_{0}$ corresponds to the
contribution from the two leading protons to the total multiplicity,
while $\Delta m$ takes the contribution of the masses of the two
participating constituent quarks to the centre-of-mass energy into
account. Having varied the $n_{0}$ and $\Delta m$ values three
different inelasticities were defined. In one of the results the
inelasticity decreases from $K$ $\sim$ (0.55 - 0.6) at the ISR
energies to 0.4 at $\sqrt{s}$ $=$ 1800 GeV. The two other results
indicated the constant value of $K$ $\sim$ 0.35.

Investigating the very high energy $pp$ interactions by cosmic ray
data it was shown \cite{Wibig} that the Feynman scaling violation,
in the form proposed by Wdowczyk and Wolfendale, leads to continuous
decrease of the inelasticity, which was found be consistent with LHC
measurements up to 7 TeV, qualitatively in agreement with our
results, Fig. (11).

In another work \cite {Navarra2003} and by using methods of
information theory approach, calculations of the inelasticity
coefficient and its energy dependence were studied, resulting that
the inelasticity remains essentially constant in energy, except for
a variation around $K$ $\sim$ 0.5 in the range 20 $<$ $\sqrt{s}$ $<$
1800 GeV to $p+p(\overline{p})$ data.

The Interacting Gluon Model (IGM) was an approach used in studies
about the inelasticities and leading particle spectra in hadronic
and nuclear collisions
\cite{YH,Navarra1993,Navarra1994,Navarra2003}. In \cite{Navarra1993}
an extended version of the IGM incorporating the production of
minijets was applied and, as a result, it was concluded that the
inelasticity slowly increases towards some limited value. The
inclusion of minijets reversed the trend of decreasing
inelasticities found in previous calculations with the IGM.

In subsequent work \cite{Navarra1994} the authors introduced a
hadronization mechanism in the IGM concluding that the minijet
production leads to inelasticities increasing with $\sqrt{s}$ and
that hadronization process does not change this trend.

Based on the above considerations, one can note that the various
approaches are largely in conflict with each other in explaining the
energy dependence of the inelasticity, reflecting the subtlety of
the theme. Hence, we have based the present study on the
experimental information on charged particle multiplicity
distributions in $p+p(\overline{p})$ collisions. Thus, we provided a
new argument in favor of the hypothesis that the $K(s)$ decreases as
a function of the center of mass energy.

\begin{table}[htb!]
\caption{\label{tabpi} Results reproduced from Ref. \cite{Beggio
Luna}, where the $P_{n}$ \emph{phenomenological procedure} was
applied.}
\begin{tabular}{ccccc}
\hline $\sqrt{s}$ GeV & $\zeta(s)$ & $\chi^{2}/N_{DF}$ & $\xi(s)$ &
$\langle N(s)\rangle
$ \\
\hline
30.4 & 0.239 $\pm$ 0.011 & $0.588$& $1.642$ & $ 9.43$ \\
44.5 & 0.240 $\pm$ 0.011 & $0.306$& $1.643$ & $10.86$ \\
52.6 & 0.239 $\pm$ 0.009 & $0.765$& $1.639$ & $11.55$ \\
62.2 & 0.231 $\pm$ 0.008 & $1.717$& $1.613$ & $12.25$ \\
300  & 0.263 $\pm$ 0.003 & $0.608$& $1.589$ & $24.47$ \\
546  & 0.305 $\pm$ 0.004 & $0.300$& $1.599$ & $29.53$ \\
1000 & 0.288 $\pm$ 0.005 & $1.469$& $1.508$ & $38.46$ \\
1800 & 0.315 $\pm$ 0.002 & $0.782$& $1.468$ & $44.82$ \\
7000 & 0.352             & $     $& $1.308$ & $81.79$ \\
14000 & 0.372            & $     $& $1.209$ & $108  $ \\\hline
\end{tabular}
\end{table}

\begin{table}[htb!]
\caption{\label{tabpi} Averaged impact parameter, $<b>$, at the ISR
energies and the corresponding inelasticity values.}
\begin{tabular}{ccccc}
\hline $\sqrt{s}$ GeV & $<b>$ & $K(s,<b>)$ \\
\hline
30.4 & 0.83 & $0.15$ \\
44.5 & 0.78 & $0.16$ \\
52.6 & 0.77 & $0.16$ \\
62.2 & 0.83 & $0.15$ \\\hline &$<K>_{ISR}$ $\sim$ 0.16\\\hline
\end{tabular}
\end{table}

\section{Concluding remarks}

In the absence of sufficient experimental information on the energy
dependence of the inelasticity to test the several existing model
predictions, we have based our analysis in the connection between
$K(s)$ and the full phase space $P_{n}(s)$ by using a satisfactory
modeling to $P_{n}(s)$ adjusted for the experimental reality over a
large range of energy, 30 $<$ $\sqrt{s}$ $\leq$ 1800 GeV, which is
consistent with several QCD prescriptions \cite{Beggio Luna}.

In the present approach $K(s,b)$ $\propto$ $\chi_{I}(s,b)$, Eq.
(\ref{eq27}), and we have adopted the DGM QCD inspired eikonal model
\cite{luna008,luna009,Beggio Luna}. The only free parameter in the
$P_{n}(s)$ formalism adjusted to $p+p(\overline{p})$ experimental
data is $\zeta(s)$, (Eq. (\ref{eq21}) - Table I), while all the
parameters of the eikonal function,
$\chi_{pp}^{\overline{p}p}(s,b)$, were determined carrying out a
global fit to $\sigma_{tot}^{pp,\overline{p}p}$,
$\rho^{pp,\overline{p}p}$ and $d\sigma_{el}^{\overline{p}p}/dt$
data. The results of the fits to $\sigma_{tot}^{pp,\overline{p}p}$,
$\rho^{pp,\overline{p}p}$ and $d\sigma_{el}^{\overline{p}p}/dt$ are
presented in \cite {Beggio Luna}. Our results predict the average
inelasticity to be $\sim$ 0.5 at the ISR energies if calculated at
$b$ $\sim$ 0, in agreement with that from Refs.
\cite{Brick1981,Kadija,Golyak} (see Section III).

The term $\chi_{gg}(s,b)$ in the Eq. (\ref{eq17}) gives the main
contribution to high multiplicities, being the responsible for the
rise of the cross sections with $\sqrt{s}$. Thus, we have concluded
that minijets from semihard interactions, arising from scattering of
gluons carrying only a very small fraction of the momenta of their
parent protons, are the responsible for the decrease of the
inelasticity as a function of the $\sqrt{s}$.

Results obtained by using alternative methods to estimate the energy
dependence of the inelasticity are in conflict with each other.
Thus, based on the experimental information on charged particle
multiplicity distributions in $p+p(\overline{p})$ collisions we
provided new evidence in favor of the hypothesis that the $K(s)$
decreases when $\sqrt{s}$ is increased.

\begin{acknowledgments}

The authors are grateful to Prof. M.J. Menon for several instructive
discussions and suggestions. This study was financed in part by the
Coordena\c{c}\~{a}o de Aperfei\c{c}oamento de Pessoal de N\'ivel
Superior - Brasil (CAPES) - Finance Code 001. We are also thankful
to the referee for valuable comments and suggestions.

\end{acknowledgments}


\newpage

\begin{figure}[bt!]
\begin{center}
\includegraphics*[height=9cm,width=8.4cm]{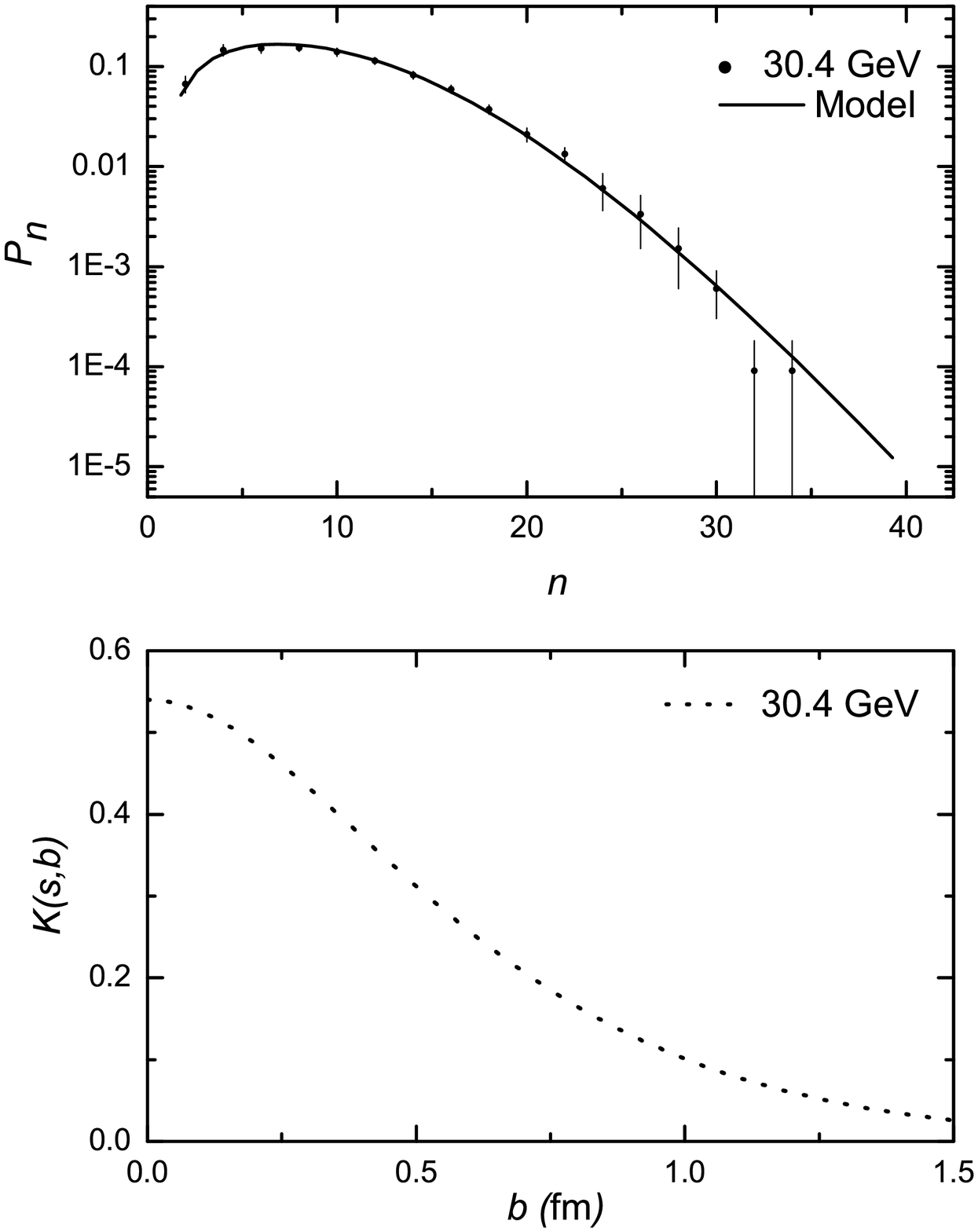}
\caption{Top panel: Comparison of the theoretical, Eqs. (\ref{eq14})
and (\ref{eq15}), and experimental results in full phase space
$P_{n}(s)$ at 30.4 GeV. Data points from \cite{ABC}. The another
panel shows prediction of $K(s)$, Eq. (\ref{eq27}), by using the
parameters obtained from $P_{n}(s)$ analysis done in \cite{Beggio
Luna}.} \label{fig01}
\end{center}
\end{figure}

\begin{figure}[ht!]
\begin{center}
\includegraphics*[height=9cm,width=8.4cm]{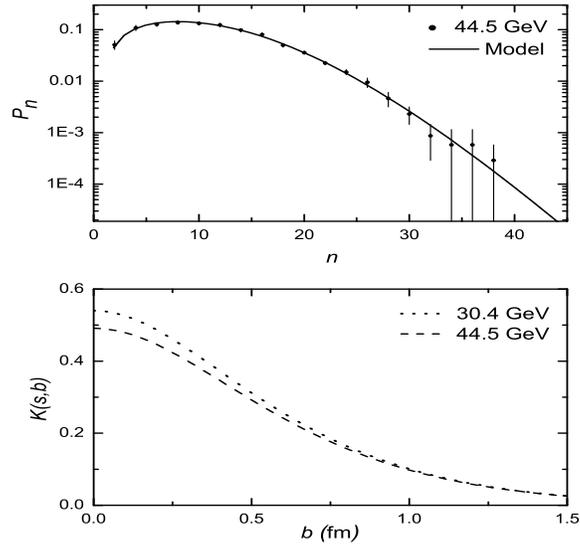}
\caption{Same as figure 1 but at 44.5 GeV.} \label{fig02}
\end{center}
\end{figure}

\begin{figure}[ht!]
\begin{center}
\includegraphics*[height=9cm,width=8.4cm]{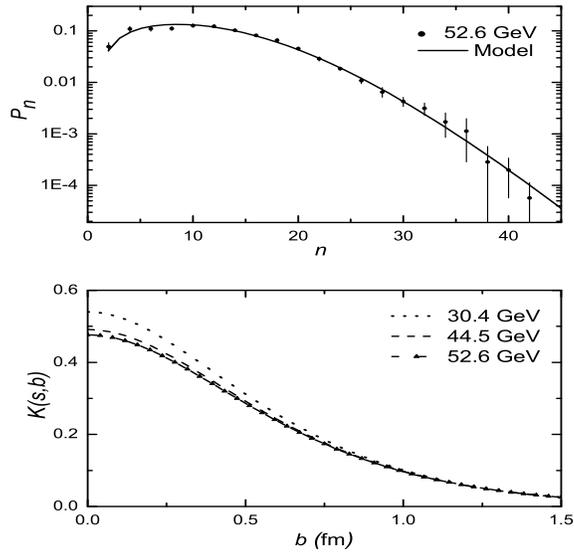}
\caption{Same as figure 1 but at 52.6 GeV.} \label{fig03}
\end{center}
\end{figure}

\begin{figure}[ht!]
\begin{center}
\includegraphics*[height=9cm,width=8.4cm]{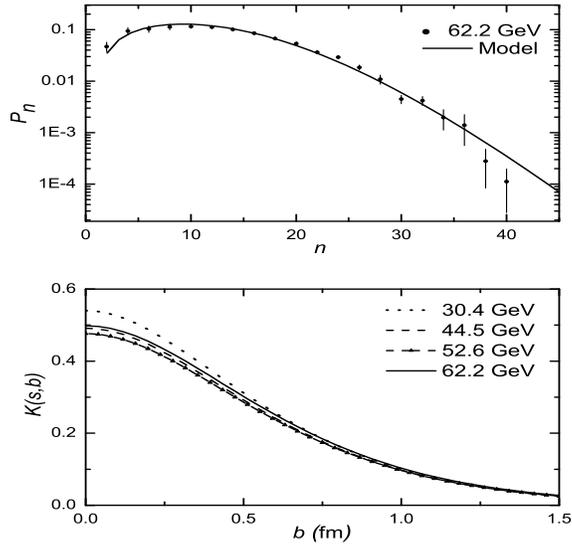}
\caption{Same as figure 1 but at 62.2 GeV.} \label{fig04}
\end{center}
\end{figure}

\begin{figure}[ht!]
\begin{center}
\includegraphics*[height=9cm,width=8.4cm]{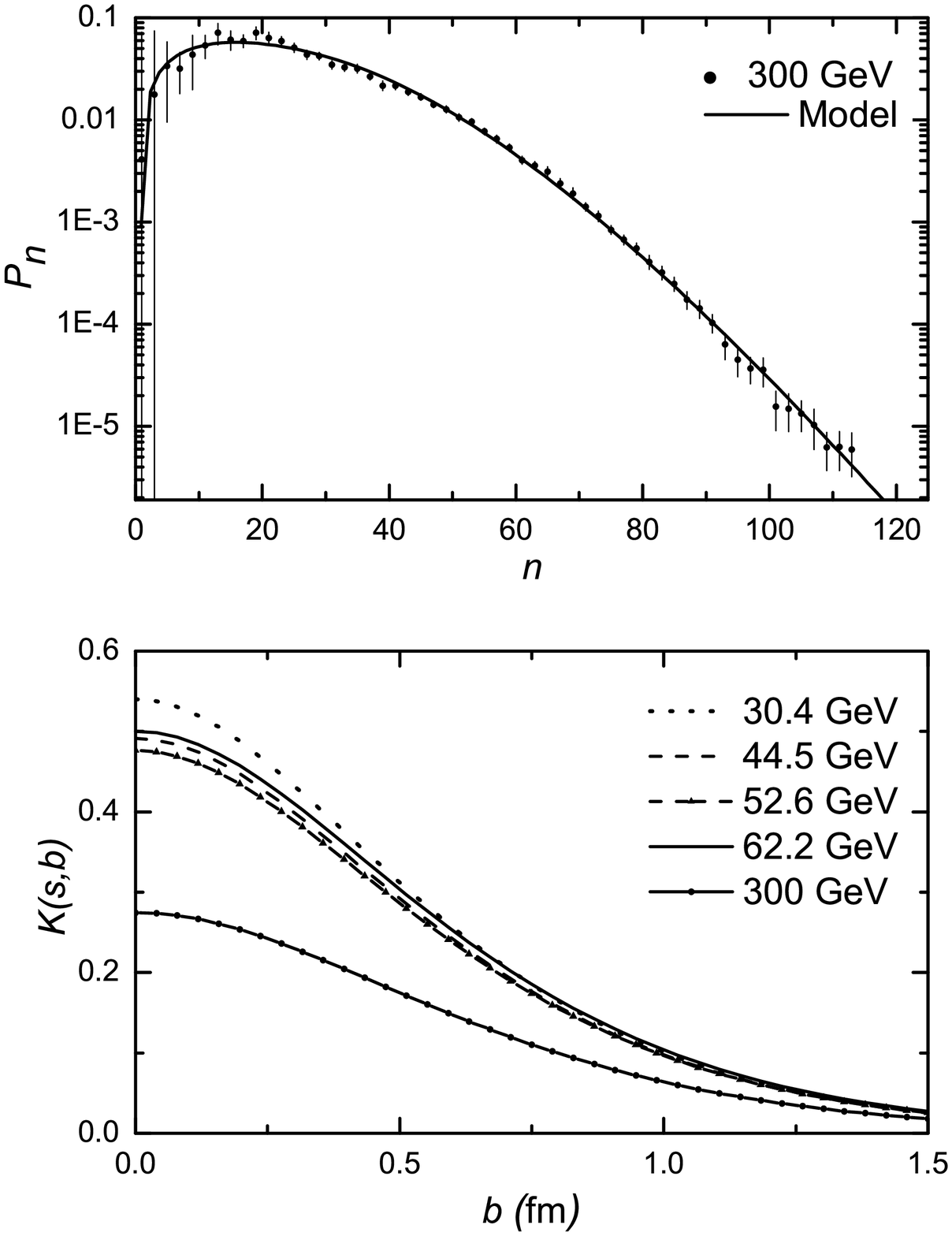}
\caption{Same as figure 1 but at 300 GeV. Data points from
\cite{Alexopoulos}.} \label{fig05}
\end{center}
\end{figure}

\begin{figure}[ht!]
\begin{center}
\includegraphics*[height=9cm,width=8.4cm]{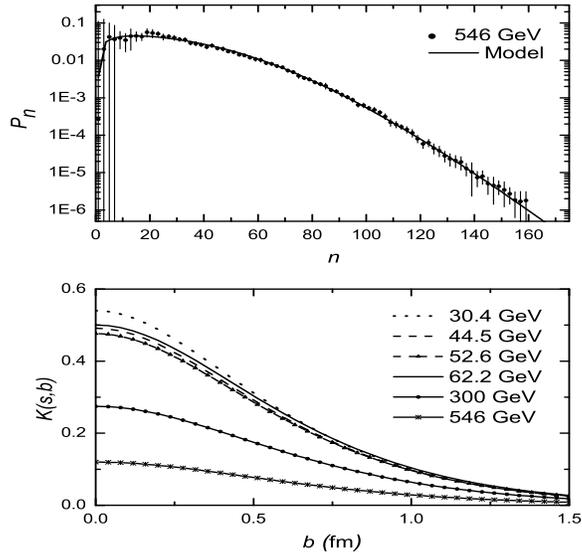}
\caption{Same as figure 1 but at 546 GeV. Data points from
\cite{Alexopoulos}.} \label{fig06}
\end{center}
\end{figure}

\begin{figure}[ht!]
\begin{center}
\includegraphics*[height=9cm,width=8.4cm]{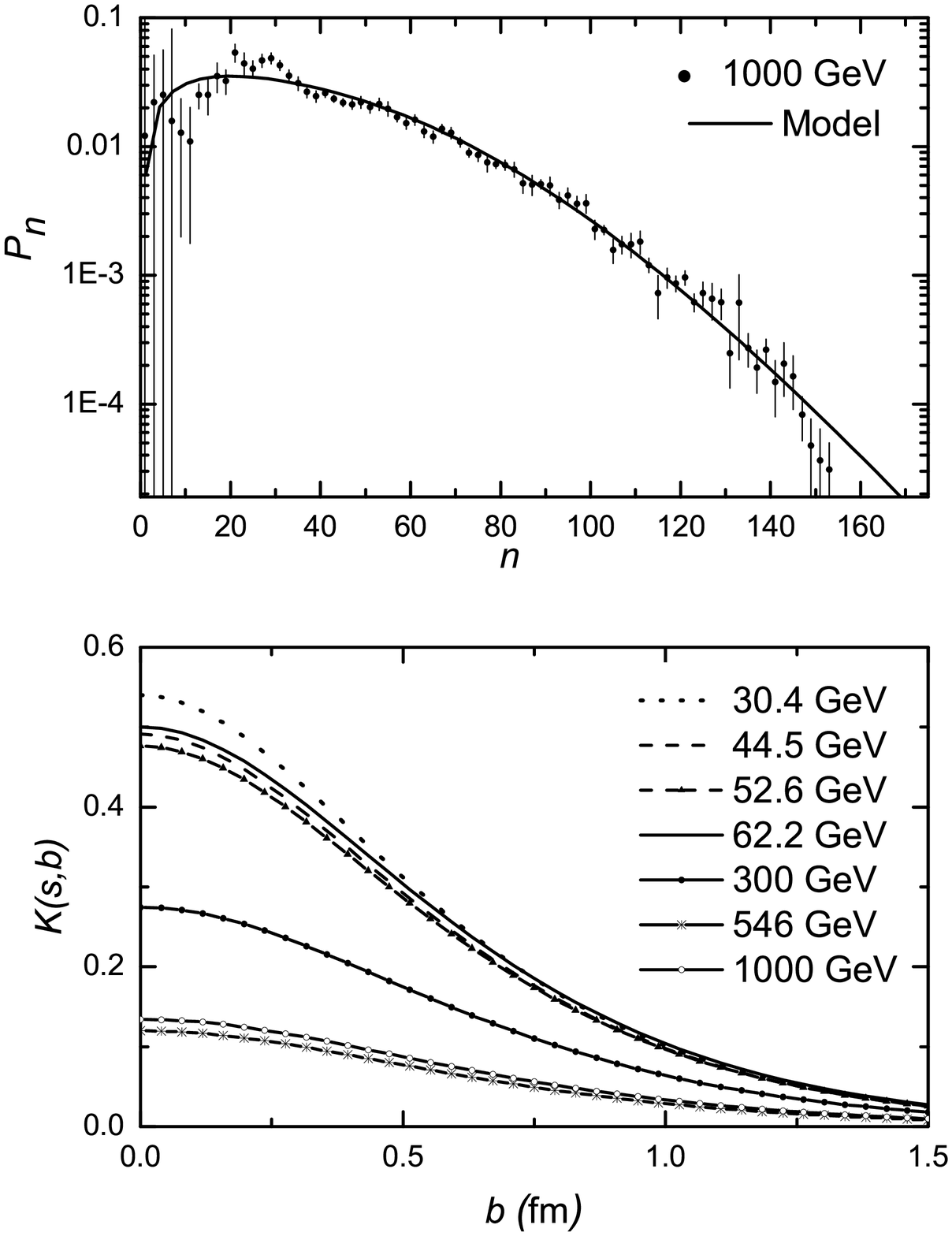}
\caption{Same as figure 1 but at 1000 GeV. Data points from
\cite{Alexopoulos}.} \label{fig07}
\end{center}
\end{figure}

\begin{figure}[ht!]
\begin{center}
\includegraphics*[height=9cm,width=8.4cm]{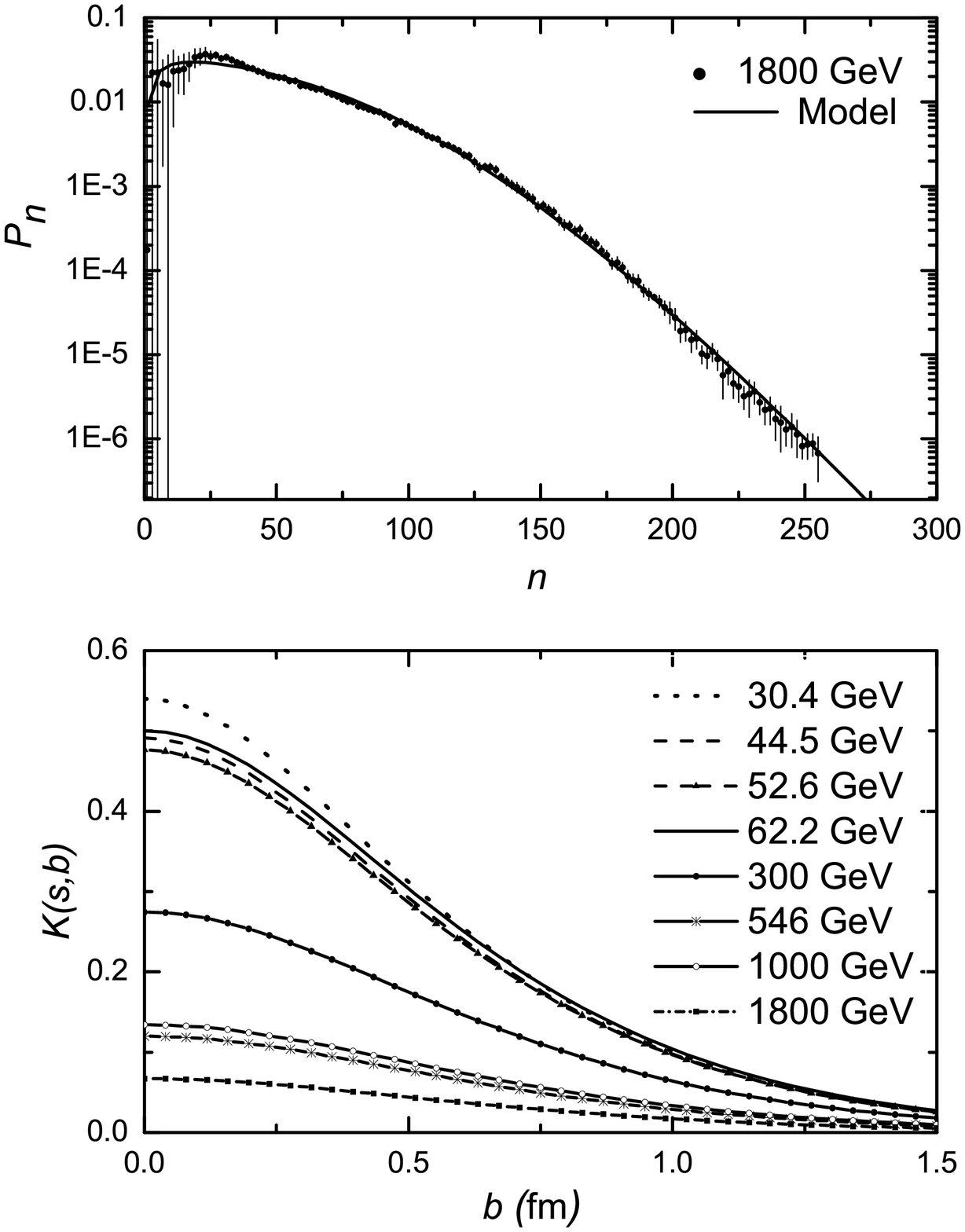}
\caption{Same as figure 1 but at 1800 GeV. Data points from
\cite{Alexopoulos}.} \label{08}
\end{center}
\end{figure}

\begin{figure}[htt!]
\begin{center}
\includegraphics*[height=9cm,width=8.4cm]{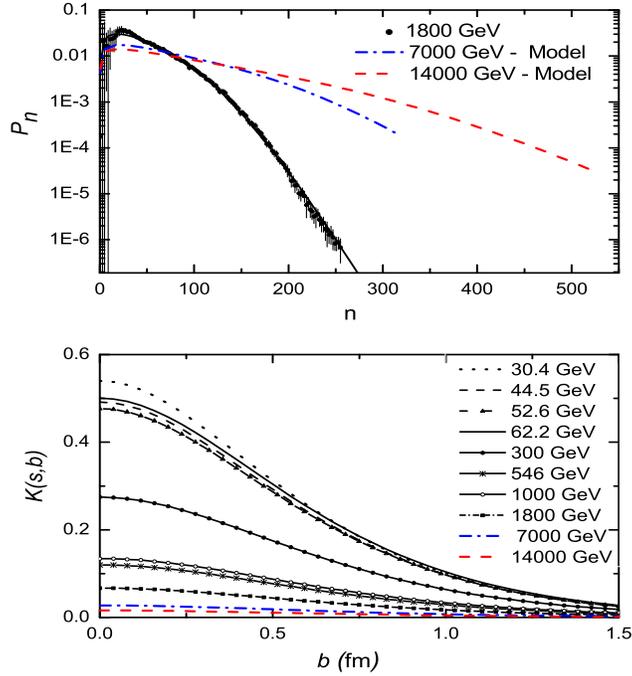}
\caption{Top panel: Theoretical results in full phase space
$P_{n}(s)$ at 7 and 14 TeV, Eqs. (15) and (16). Experimental data of
the $P_{n}(s)$ at 1800 GeV added to comparison. The another panel
shows predictions of $K(s)$, Eq. (\ref{eq27}), by using parameters
obtained from $P_{n}(s)$ analysis done in \cite{Beggio Luna}.}
\label{fig09}
\end{center}
\end{figure}

\begin{figure}[htt!]
\begin{center}
\includegraphics*[height=9cm,width=8.4cm]{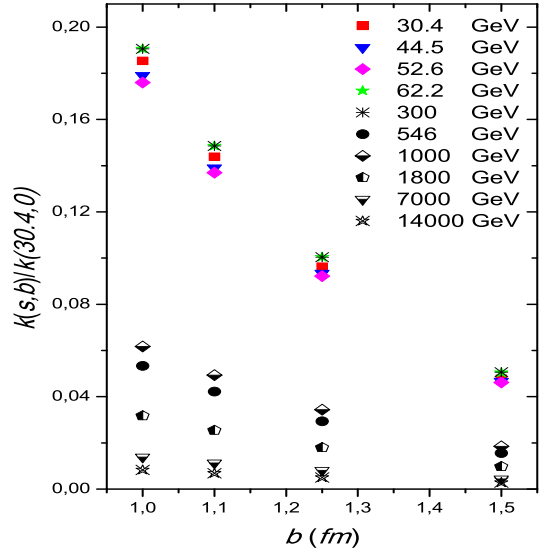}
\caption{Ratios $K(s,b)/K(30.4,0)$ calculated for different collison
energies and impact parameters values. The value of $K(30.4,0)$ is
0.54, Fig. (1). The particle production processes tend to be more
peripheral at the ISR specific energies of 30.4, 44.5, 52.6 and 62.2
GeV.} \label{fig10}
\end{center}
\end{figure}

\begin{figure}[htt!]
\begin{center}
\includegraphics*[height=9cm,width=8.4cm]{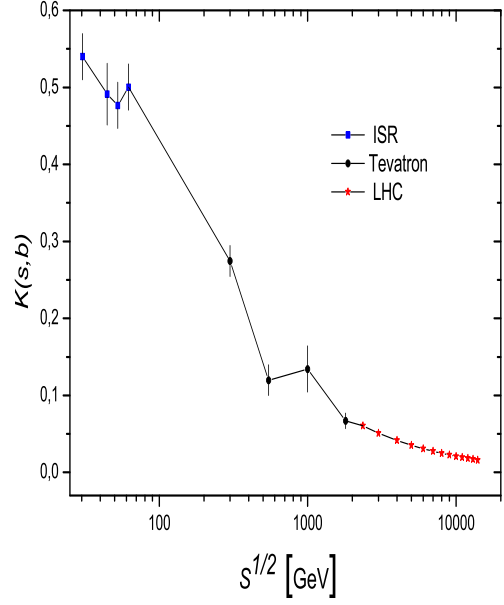}
\caption{Inelasticities calculated at $b$ $\approx$ $0$ as a
function of center mass energy, Eq. (\ref{eq27}). The results show
marked decrease in the inelasticity from ISR to LHC energies. The
error bars represent the uncertainties of the parameters $\gamma$
and $\zeta$ propagated to the inelasticity values, the line is drawn
only as guidance for the points.} \label{fig11}
\end{center}
\end{figure}

\end{document}